\documentclass[a4paper,10pt,aps,prl,twocolumn,showpacs,amsmath,amssymb, floatfix]{revtex4}
\usepackage{graphicx,color}

\newcommand{\dd}{\mathrm{d}}

\newcommand{\mean}[1]{\langle #1 \rangle}

\newcommand{\IInt}[3]{\int_{#2}^{#3}\dd #1\;}

\renewcommand{\vec}[1]{\mathbf #1}

\newcommand{\al}{\alpha}
\newcommand{\gam}{\gamma}

\newcommand{\vhi}{\varphi}

\newcommand{\x}{\vec r}

\newcommand{\tr}{\tau_\text{r}}
\newcommand{\Dc}{D_\text{c}}

\begin{document}

\title{Collective behavior of quorum-sensing run-and-tumble particles in
  confinement}

\author{Markus Rein}
\author{Nike Hein\ss}
\author{Friederike Schmid}
\author{Thomas Speck}
\affiliation{Institut f\"ur Physik, Johannes Gutenberg-Universit\"at Mainz,
  Staudingerweg 7-9, 55128 Mainz, Germany}

\begin{abstract}
  We study a generic model for quorum-sensing bacteria in circular
  confinement. Every bacterium produces signaling molecules, the local
  concentration of which triggers a response when a certain threshold is
  reached. If this response lowers the motility then an aggregation of
  bacteria occurs, which differs fundamentally from standard motility-induced
  phase separation due to the long-ranged nature of the concentration of
  signal molecules. We analyze this phenomenon analytically and by numerical
  simulations employing two different protocols leading to stationary cluster
  and ring morphologies, respectively.
\end{abstract}



\maketitle


\textit{Introduction.} Motility and locomotion are basic as well as
challenging tasks for microorganisms exploring complex aqueous
environments~\cite{laug09}, and nature has developed a range of diverse
strategies for this purpose. For example, the sperm cells of sea urchins find
the egg by moving along helical paths, the curvature of which is controlled by
the concentration of a chemoattractant~\cite{frie07,jike15}. On the other
hand, bacteria might use quorum sensing to respond to changes in their
environment~\cite{mill01}. The probably most famous example is the marine
bacterium \emph{V. fischeri}, which controls bioluminescence in accordance
with its population density. To this end bacteria measure the local
concentration of certain signal molecules, called autoinducers, which are
emitted by other bacteria.

Moving along a chemical gradient of chemoattractants or repellents is called
chemotaxis. The arguably most famous and best studied model for chemotaxis is
the Keller-Segel model~\cite{kell70,bren98}, which consists of two coupled
partial differential equations, one for the density of diffusing bacteria and
one for the concentration of signal molecules. The Keller-Segel model has
become a cornerstone to study pattern formation (such as rings and spots in
\emph{E. coli}~\cite{adle66,budr91} and \emph{S. typhimurium}~\cite{wood95})
and self-organization in general~\cite{hill09}. It is not restricted to
bacteria, \emph{e.g.}, chemotactic behavior has also been reported for
self-propelled colloidal particles~\cite{paxt04,theu12,pohl14}, which are
phoretically driven by the catalytic decomposition of, \emph{e.g.}, hydrogen
peroxide playing the role of the chemical signal.

It has been argued that chemotaxis is not the only route to self-organization
of motile cells and bacteria, and that similar patterns are observed in an
arrested motility-induced phase transition in combination with bacteria
reproduction~\cite{cate10,cate12}. Such a scenario relies on a positive
feedback through a density-dependent motility with ``slow'' bacteria in dense
environments~\cite{cate15} so that they may move against a density
gradient. It allows an effective equilibrium description in terms of a
coarse-grained population density as long as the motility is a local function,
which seems to be a good assumption for short-ranged physical interactions,
\emph{e.g.}, for self-propelled colloidal
particles~\cite{butt13,spec14,spec15}. However, for quorum-sensing bacteria
the motility is no longer a function of the density but of the local
concentration of the autoinducers. The dependence of the local concentration
on the sources (or sinks in the case of catalytic swimmers) is strongly
non-local and long-ranged, which precludes a mapping onto equilibrium phase
separation.

In this Letter we study how patterns can emerge based on motility changes even
in populations with a conserved number of members. To this end, we combine a
simple model for bacteria dynamics with quorum sensing. Bacteria (or more
generally, particles) solely interact via signaling molecules. Particles are
confined and we observe aggregation in the center of the confinement mediated
by the autoinducers (see Refs.~\citenum{park03,mars14} for experiments and
numerical results in more complex confining geometries). This aggregation is
in contrast to other collective behavior of confined active particles like the
self-organized pump in a harmonic trap~\cite{nash10,henn14} and the
aggregation at walls~\cite{fily14a,yang14,smal15}. By combining numerical
simulations and analytical theory, we show that the aggregation is determined
by a set of universal parameters that depend on system size.


\textit{Model.} We model the bacteria as run-and-tumble particles moving in
two dimensions above a substrate. The dynamics mimics straight ``runs'' due
to, \emph{e.g.}, synchronized flagella interrupted by random ``tumble''
events~\cite{poli09,reig12}. The equations of motion are
\begin{equation}
  \label{eq:dyn}
  \dot\x_k = v\vec e_k + \mu_0\vec F_k,
\end{equation}
where $\vec F_k$ is the force and $\mu_0$ the bare mobility. Every particle
has an orientation $\vec e_k=(\cos\vhi_k,\sin\vhi_k)^T$ along which it is
propelled with speed $v$. This orientation remains fixed for an exponentially
distributed random waiting time with mean $\tr$, after which a tumble event
occurs. We assume the tumbling to occur instantaneously and pick a new,
uniformly distributed, orientation $\vhi_k$~\cite{schn93}.

Every particle produces autoinducers with rate $\gam$. These signal molecules
with concentration $\hat c(\x,t)$ diffuse with diffusion coefficient
$\Dc$. While the actual particles move in two dimensions and are confined by a
circular confinement with radius $R$ (\emph{e.g.}, due to a semi-permeable
membrane), the autoinducers can penetrate this wall and permeate the semispace
above the substrate. The time evolution of the concentration is thus described
by
\begin{equation}
  \label{eq:c:diff}
  \partial_t\hat c = \Dc\nabla^2\hat c + \gam\sum_{k=1}^N\delta(\x-\x_k).
\end{equation}
In the following we assume that the molecular diffusion is much faster than
the motion of the much larger particles so that there is an instantaneous
stationary concentration of autoinducers
\begin{equation}
  \hat c(\x) = \frac{\gam}{4\pi\Dc}\sum_{k=1}^N\frac{1}{|\x-\x_k|}.
\end{equation}
Due to the autoinducers eventually leaving the confinement the concentration
remains finite. The collective behavior of the particles is controlled through
$\tr(\x)=\tr(\hat c(\x))$ and $v(\x)=v(\hat c(\x))$, which both can vary
spatially through their dependence on the concentration of autoinducers
$c(\x)$. Averaging over particle positions gives the average concentration
$c(\x)=\mean{\hat c(\x)}$.

\begin{figure}[t]
  \centering
  \includegraphics[width=.9\linewidth]{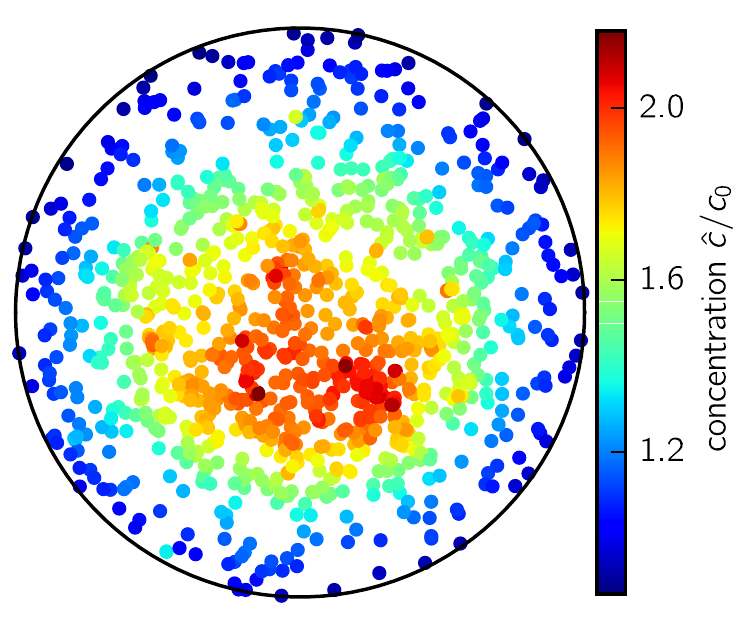}
  \caption{Snapshot of a configuration with $N=1000$ quorum-sensing
    run-and-tumble particles. The color code indicates the local concentration
    $\hat c$ of signal molecules felt by each particle.}
  \label{fig:snap}
\end{figure}

For the numerical integration of Eq.~\eqref{eq:dyn} we employ a fixed time
step $\delta t$. After propagating all particles along their orientation, a
new random orientation is assigned with probability $\delta t/\tr$. Instead of
an ideal hard wall, we employ the Weeks-Chandler-Andersen (WCA)
potential~\cite{week72} leading to a small but finite ``thickness''
$d=5\cdot10^{-3}R$ of the wall. Particles with outward-pointing orientations
remain trapped within the wall until the orientations, due to their rotational
diffusion, point inwards again. Fig.~\ref{fig:snap} shows a snapshot of the
system for $N=1000$ particles after relaxation to the steady state~[SM].


\textit{Mean-field theory.} We first consider run-and-tumble particles that
only interact via sensing the autoinducers. It is then sufficient to consider
the one-point density $\psi(\x,\vhi,t)$ of position and orientation, which
obeys the dynamical equation
\begin{equation}
  \label{eq:psi}
  \partial_t\psi = -\nabla\cdot(v\vec e\psi) -
  \frac{1}{\tr}\psi + \frac{1}{2\pi\tr}\rho
\end{equation}
with particle density $\rho(\x,t)=\IInt{\vhi}{0}{2\pi}\psi(\x,\vhi,t)$
corresponding to the zeroth moment of $\psi$. The first moment
$\vec p(\x,t)=\IInt{\vhi}{0}{2\pi}\vec e\psi(\x,\vhi,t)$ describes the
orientational density of the run-and-tumble particles. From Eq.~\eqref{eq:psi}
we obtain the adiabatic solution $\vec p=-\frac{1}{2}\tr\nabla(v\rho)$
dropping the time derivative and neglecting the dependence on the second
moment.

For the analytical treatment we further approximate $v=v(c)$ and $\tr=\tr(c)$,
\emph{i.e.}, the response depends on the local \emph{average} concentration
$c(\x)$. It is instructive to first consider the Keller-Segel model, which,
assuming constant $\tr$, follows in the limit of a weak perturbation of the
velocity $v(c)\approx\bar v+v'(\bar c)(c-\bar c)$ around a uniform
concentration $\bar c$. Eq.~\eqref{eq:psi} implies
$\partial_t\rho=-\nabla\cdot(v\vec p)$, which leads to
\begin{equation}
  \partial_t\rho = \nabla\cdot(D\nabla\rho - \chi\rho\nabla c)
\end{equation}
after inserting the adiabatic solution for the orientational density. This is
the Keller-Segel model together with
$\partial_tc=\Dc\nabla^2c+\gam\rho\delta(z)$, where the source term in
Eq.~\eqref{eq:c:diff} has been replaced by the density
$\rho_\text{3D}(\x,z)=\rho(\x)\delta(z)$. The two coefficients
$D=\frac{1}{2}\tr\bar v^2$ and $\chi=-\frac{1}{2}\tr\bar v v'(\bar c)$ are the
effective diffusivity and chemotactic sensitivity, respectively. This result
demonstrates that chemotactic behavior can be achieved simply by changing the
magnitude of the speed depending on the difference between the local and a
fixed reference concentration without sensing the concentration gradient.

\begin{figure*}[t]
  \centering
  \includegraphics[width=\linewidth]{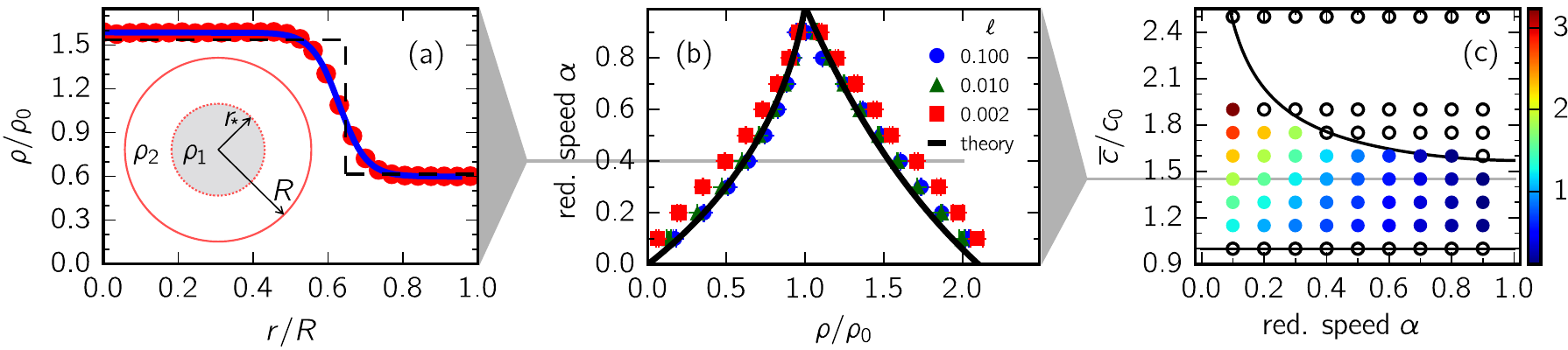}
  \caption{Clustering of slow particles. (a)~The particle density $\rho(r)$
    for reduced speed $\al=0.4$ and persistence length $\ell=0.01$ obtained
    numerically for $N=1000$ particles (symbols) and from the mean-field
    theory (dashed line). The solid line is a fit to Eq.~(\ref{eq:rho}). A
    sketch of a cluster with radius $r_\ast$ and density $\rho_1$ is shown in
    the inset, where $\rho_2$ is the density of the dilute region surrounding
    the cluster. (b)~Densities $\rho_1$ (dense, right branch) and $\rho_2$
    (dilute, left branch) for threshold $\bar c=1.45c_0$. The numerical
    results slightly depend on the persistence length $\ell$, still the
    overall agreement with the theoretical prediction is
    excellent. (c)~Numerical phase diagram of reduced speed $\al$ \emph{vs.}
    threshold $\bar c$ for $\ell=0.01$. The color bar indicates the density
    difference $(\rho_1-\rho_2)/\rho_0$, whereas open circles indicate that no
    formation of a cluster has occurred. Also shown is the theoretical
    prediction (solid lines).}
  \label{fig:circ}
\end{figure*}

In the following, however, we are rather interested in collective effects that
arise because of large (discontinuous) changes of the speed due to some
particles reaching a threshold, which is thus beyond the scope of the
Keller-Segel model. In the steady state, $\partial_t\psi=0$ and we obtain from
Eq.~\eqref{eq:psi}
\begin{equation}
  \label{eq:ss}
  \vec p = -\frac{\tr}{2}\nabla(v\rho), \qquad 0 = \nabla\cdot(v\vec p),
\end{equation}
where we have neglected the dependence on the second moment. For simplicity,
the confinement is now modeled through the no-flux boundary condition
$\vec n\cdot\vec p|_R=0$ with wall normal $\vec n$, which ignores the trapping
of particles at the wall due to their persistent motion. To compare the theory
with the numerical results, we define the effective bulk density
$\rho_0=N_\text{bulk}/(\pi R^2)$ with $N_\text{bulk}$ the average number of
run-and-tumble particles inside $r<R$ determined in the simulations.

Exploiting the mentioned time-scale separation between molecular diffusion and
particle motion, the concentration of autoinducers follows from
Eq.~\eqref{eq:c:diff} with $\partial_tc=0$ and the source term again replaced
by the density $\rho_\text{3D}(\x,z)=\rho(\x)\delta(z)$,
$\nabla^2c(\x)=-(\gam/\Dc)\rho(\x)\delta(z)$, which is Poisson's
equation. Since the concentration $c(\x)$ determines the speed via $v(c)$ it
is coupled with Eq.~\eqref{eq:ss}, which can now be solved for the density
profile $\rho(r)$. In the remainder of this Letter we will discuss the two
situations of one and two thresholds.


\textit{Piecewise constant speed.} We now specialize to a piecewise constant
speed $v(c)$. Suppose that there are two regions with different speeds. Within
each region we find $\nabla\cdot\vec p=0$ from Eq.~\eqref{eq:ss} and hence the
normal components of $\vec p$ across the interface have to be equal. For
non-vanishing $\vec p$ this would imply a steady particle current, which is
excluded by the no-flux boundary condition. Hence, we conclude that $\vec p=0$
and, within our theory neglecting fluctuations,
$v\rho=\text{const}$. Interestingly, the reorientation time $\tr$ drops out
and does not influence the steady state. This is quite in contrast to
self-propelled particles with volume exclusion, the collective behavior of
which is strongly influenced by orientation relaxation~\cite{cate15,spec15}.

Specifically, we introduce a threshold concentration $\bar c$ above which the
particles slow down by a factor $\al\leqslant 1$,
\begin{equation}
  v(\hat c) = 
  \begin{cases}
    v_0 & (\hat c < \bar c), \\ v_0\al & (\hat c \geqslant \bar c)
  \end{cases}
\end{equation}
with reference speed $v_0$. In the following, the importance of the directed
motion is captured by the persistence length $\ell=v_0\tr/R$ divided by the
radius of the confinement. Due to the slow decay of $\hat c(\x)$, we expect
that the steady state will have a radial symmetry with an inner dense cluster
and a dilute outer region. In Fig.~\ref{fig:circ}(a) the density profiles
predicted by the theory and measured in the simulations are shown for
$\al=0.4$ and $N=1000$ particles. It clearly shows a higher inner density
corresponding to the cluster and a lower outer density.

The theoretical density is a step function that can be obtained as
follows. The radius $r_\ast$ of the cluster is determined by the condition
$c(r_\ast)=\bar c$. Since $v\rho=\text{const}$, the density in both regions is
constant with dilute density $\rho_2=\al\rho_1$. Taking into account the
conservation of the bulk density
\begin{equation}
  \rho_0 = \frac{1}{\pi R^2}\IInt{r}{0}{R} 2\pi r\rho(r),
\end{equation}
the density of the cluster follows as
\begin{equation}
  \rho_1 = \frac{\rho_0}{\al+(1-\al)x_\ast^2}
\end{equation}
with $x_\ast=r_\ast/R$. The radially symmetric solution of the Poisson
equation reads
\begin{equation}
  c(r) = \frac{\gam}{\pi\Dc}\IInt{r'}{0}{R}\mathcal K(r'/r)\rho(r')
\end{equation}
with kernel $\mathcal K(x)=xK(x^2)$ for $x<1$ and $\mathcal K(x)=K(x^{-2})$
for $x>1$, where $K(x)$ is the complete elliptic integral of the first
kind. Inserting the step profile $\rho(r)$ we determine self-consistently the
radius $r_\ast$ of the cluster and thus $\rho_1$ and $\rho_2$ for given
$\bar c$, $\al$, and $\rho_0$. There is a lower bound $c_0<\bar c$ to the
threshold below which no clustering is possible. It is obtained by considering
a homogenous density $\rho_1=\rho_0$ with interface at $r_\ast=R$ and thus
$c_0=c(R)=\gam\rho_0R/(\pi\Dc)$.

In contrast to the theoretical step profiles, the numerical profiles show a
finite interfacial width due to fluctuations. Despite the fact that these
fluctuations are not accounted for in the theory, the densities of the cluster
and the dilute outer region are predicted quite accurately (at least for not
too small persistence lengths). The full profile is well fitted by the
empirical expression
\begin{equation}
  \label{eq:rho}
  \rho(r) = \frac{\rho_1+\rho_2}{2} -
  \frac{\rho_1-\rho_2}{2}\tanh\left(\frac{r-r_\ast'}{2w}\right),
\end{equation}
from which the densities $\rho_{1,2}$ and the width $w$ of the profile can be
extracted. Fitted positions $r_\ast'<r_\ast$ of the interface are smaller than
the prediction $r_\ast$. In Fig.~\ref{fig:circ}(b) the densities $\rho_{1,2}$
are plotted as a function of reduced speed $\al$ for fixed threshold
concentration $\bar c=1.45c_0$ and several values of the persistence length,
which show good agreement with the theoretical curve.

The phase diagram in the $\al$--$\bar c$ plane is shown in
Fig.~\ref{fig:circ}(c). Besides the lower threshold $c_0$ there is also an
upper threshold depending on $\al$ beyond which no clustering is possible
anymore. After a bit of algebra one finds for the concentration at the
interface
\begin{equation}
  \frac{c_\ast}{c_0} =
  \frac{\al E(x_\ast^2)+(1-\al)x_\ast}{\al+(1-\al)x_\ast^2}
\end{equation}
with $E(x)$ the complete elliptic integral of the second kind. This function
has a maximum for $0\leqslant x_\ast\leqslant 1$, which means that a threshold
higher than this maximum cannot be reached and, therefore, no clustering is
possible. Again, the theoretical predictions for the parameter space where
clustering is possible agree very well with the numerical observations as
shown in Fig.~\ref{fig:circ}(c).


\begin{figure}[t]
  \centering
  \includegraphics{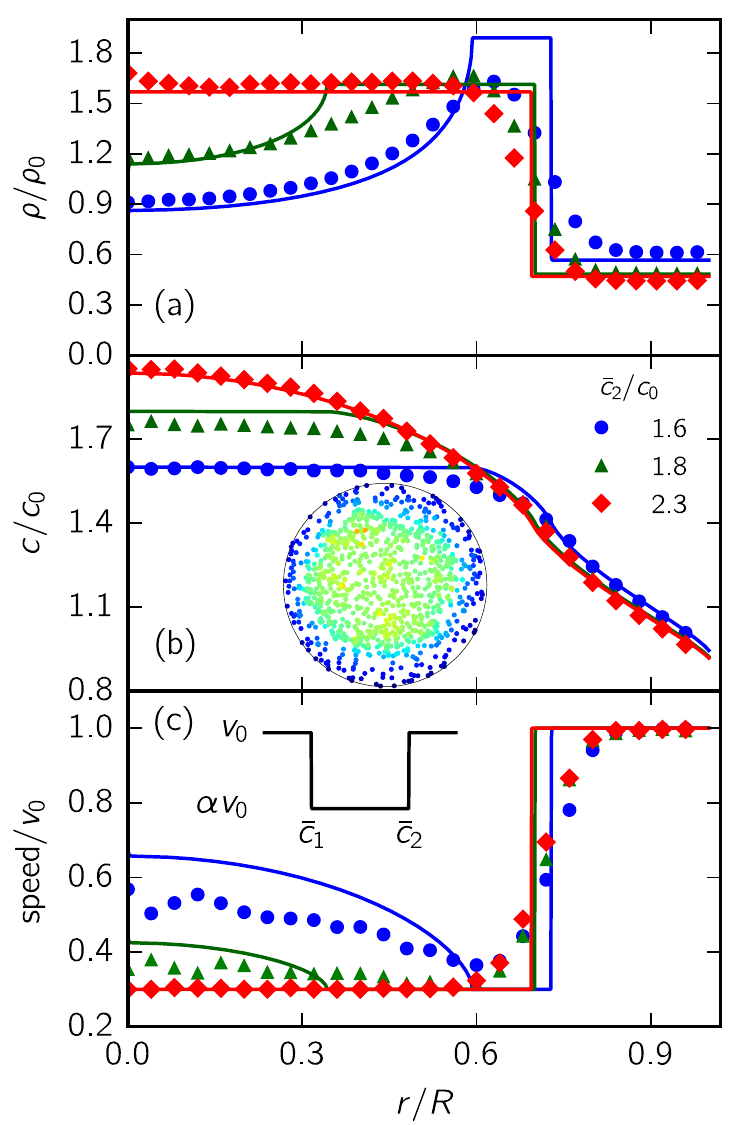}
  \caption{Formation of a ring for two thresholds $\bar c_2>\bar c_1$.  Shown
    is (a)~the particle densities $\rho(r)$, (b)~the concentration profiles
    $c(r)$, and (c)~the actual average speeds $\mean{v(\hat c)}$ as functions
    of the radial distance $r/R$ for three different values of the upper
    threshold $\bar c_2$ with the lower threshold $\bar c_1=1.4c_0$ held
    fixed. Symbols indicate simulation results for $N=1000$ particles, solid
    lines correspond to the mean-field predictions. Parameters are:
    persistence length $\ell=0.01$ and reduced speed $\al=0.3$. The inset in
    (b) shows a snapshot of the ring ($\bar c_2=1.6c_0$, colors as in
    Fig.~\ref{fig:snap}), in the inset of (c) the protocol is sketched.}
  \label{fig:ring}
\end{figure}

\textit{Rings.} More complicated morphologies can also be realized. To this
end we study numerically the effect a second threshold $\bar c_2>\bar c_1$ has
on the clustering, where for concentrations $\hat c\geqslant\bar c_2$ the
particles again move with the higher speed $v_0$. The resulting profiles for
density $\rho(r)$, concentration $c(r)$, and actual speed $\mean{v(\hat c)}$
are shown in Fig.~\ref{fig:ring} for a fixed first threshold
$\bar c_1=1.4c_0$. If $\bar c_2$ lies outside the region indicated in
Fig.~\ref{fig:circ}(c) where clustering is possible then basically no change
is observed. If the second threshold lies within the clustering region the
inner part of the cluster is depleted, leading to the formation of a
stationary dense ring~[SM]. Such rings have been observed, \emph{e.g.}, for
\emph{E. coli}~\cite{adle66} but have been attributed to metabolizing the
chemoattractant.

Fig.~\ref{fig:ring}(b) shows the corresponding average concentrations $c(r)$
of the autoinducers. While these are roughly independent of $\bar c_2$ outside
the ring, the concentrations saturate at the second threshold $\bar c_2$
inside the ring. Particles that locally cross the threshold move faster so
that there is an effective ``pressure'' to reduce the inner density. Indeed,
Fig.~\ref{fig:ring}(c) shows that the measured average speed
$\mean{v(\hat c)}$ is higher than $\al v_0$ in the inner region, dropping with
increasing $r$. The density maximum coincides with the minimum of the speed
before the speed again increases going towards the dilute region. While we do
not have closed analytical expressions, we can still solve the mean-field
equations numerically, whereby the speed varies between $v_0\al$ and $v_0$ and
the density follows such that the product $v\rho$ remains constant. As shown
in Fig.~\ref{fig:ring}, the mean-field solution correctly captures the
qualitative behavior with an inner region where $v>v_0\al$, a ring (of finite
width) within which $v=v_0\al$, and a sharp interface to the dilute outer
region.


\textit{Conclusions.} To conclude, we have presented a quantitative theory for
the collective behavior of quorum-sensing run-and-tumble particles. For one
threshold we have derived specific expressions for the cluster morphology in
circular confinement and we have confirmed these theoretical predictions in
numerical simulations. For a second threshold we have found the formation of a
ring, which is also correctly described by the mean-field theory. Patterns
typically ascribed to chemotaxis~\cite{budr91,wood95} or motility-induced
phase separation~\cite{cate10} could thus also be the result of a
quorum-sensing mechanism that changes the motility of single microorganisms in
response to environmental changes. In contrast to (effective) equilibrium
phase separation, the densities $\rho_{1,2}\propto\rho_0$ are proportional to
the global density. Moreover, the lower threshold $c_0\propto R$ depends on
the system size $R$ due to the long-ranged concentration profile of the
signaling molecules, which implies that clustering in a sufficiently large
system is suppressed. As a first step we have considered the most basic
combination of quorum sensing with a simplified model of directed motion. It
will be interesting to explore other morphologies and to study the basic
mechanism for aggregation in more realistic models and test the validity of
the scenario we have found.


\acknowledgments

TS acknowledges financial support by the DFG within priority program SPP 1726
(grant number SP 1382/3-1). We thank ZDV Mainz for computing time on MOGON.


\end{document}